# SADN: LEARNED LIGHT FIELD IMAGE COMPRESSION WITH SPATIAL-ANGULAR DECORRELATION

*Kedeng Tong*, *Xin Jin*, *Senior Member*, *IEEE*, *Chen Wang, Fan Jiang*

Shenzhen Key Lab of Broadband Network and Multimedia,
Shenzhen International Graduate School, Tsinghua University, Shenzhen 518055, China

## ABSTRACT

Light field image becomes one of the most promising media types for immersive video applications. In this paper, we propose a novel end-to-end spatial-angular-decorrelated network (SADN) for high-efficiency light field image compression. Different from the existing methods that exploit either spatial or angular consistency in the light field image, SADN decouples the angular and spatial information by dilation convolution and stride convolution in spatial-angular interaction, and performs feature fusion to compress spatial and angular information jointly. To train a stable and robust algorithm, a large-scale dataset consisting of 7549 light field images is proposed and built. The proposed method provides 2.137 times and 2.849 times higher compression efficiency relative to H.266/VVC and H.265/HEVC inter coding, respectively. It also outperforms the end-to-end image compression networks by an average of 79.6% bitrate saving with much higher subjective quality and light field consistency.

***Index Terms***— Light field image compression, end-to-end learned method, angular-spatial decoupling

## 1. INTRODUCTION

Light field (LF) images that record the angular and spatial information of light rays simultaneously facilitate depth estimation [1], 3D reconstruction [2] and free viewpoint rendering [3], which become one of the most promising media types for immersive applications like 6-degree-of-freedom virtual reality (6DoF VR). Thus, efficient LF image compression is under investigation in Joint Photographic Experts Group (JPEG) in JPEG PLENO [4] and ISO/IEC JTC 1/SC 29 WG04 under MPEG Video Coding [5].

Generally, a light field image can be represented by: a lenslet image (LI) as that shown in Fig. 1 (a), which consists of spatially arranged micro-images (MIs) recording the angular information at spatial positions; and sub-aperture images (SAIs) as that shown in Fig. 1 (b), each of which records spatial information at a perspective. Although an LI is generally captured by plenoptic cameras [6] (e.g. Raytrix cameras [7]) and SAIs are acquired by camera arrays, SAIs at a time can be converted to an LI by extracting the pixels at the same spatial coordinate in all the SAIs to form an MI in the LI [8, 9].

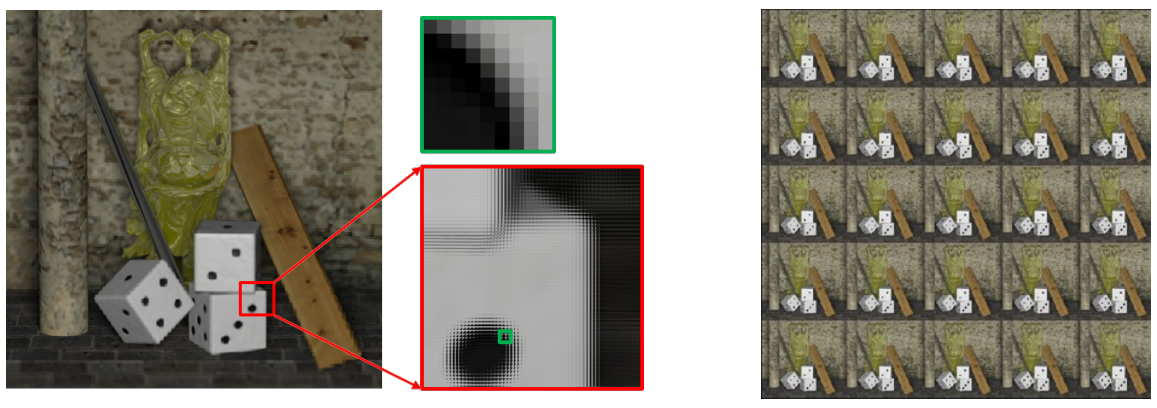

(a) (b)

**Fig. 1.** Two light field representations (a) LI, (b) SAIs.

Corresponding to the two representations, the existing LF compression methods are classified into two categories: compressing LIs [10-13] and compressing SAIs [14-19]. LI-compressing methods [10-12] exploit repetitive structures in the adjacent MIs, while providing reduced coding efficiency in the depth varying area. SAIs-compressing methods reorder [17] or sparsely sample [14] the SAIs to further exploit the angular correlations by inter coding. However, both of them concentrate on spatial consistency or angular consistency and inevitably destroy intrinsic consistency. Thus, some end-to-end networks arise to exploit the intrinsic consistency of light field based on nonlinear approximation capability. Z. Zhao *et al* [20] introduced a convolutional neural network (CNN) to reconstruct unsampled SAIs. T. Zhong *et al* [21] proposed an adaptive 3D CNN to compress the rearranged SAIs sequence along the spatial or angular dimension. However, they ignore the impact of the spatial and angular coupling in the LF image, especially that in the LI. Although there exists some well-performed hierarchical networks for end-to-end image compression [22, 23], none of them can perform well in LI compression because of lacking the solutions in decoupling the spatial and angular information.

Consequently, a novel end-to-end spatial-angular-decorrelated network (SADN) is proposed in this paper to compress the spatial and angular information in the LF image jointly. Specifically, spatial-angular interaction decouples the spatial and angular information of focused and depth varying objects from the distinctive pixel distribution in the LI by dilation convolution and stride convolution. Feature fusion fuses the spatial feature maps and angular feature maps to compress the spatial and angular information jointly. To train a robust and stable algorithm, a large-scale dataset consisting of 7549 natural light field images with diverse textures and contents is proposed and built. On average, the proposed method achieves the best performance at each rate-distortion (RD) points, and saves 64.6% bitrate relative to H.265/HEVC inter coding and 79.6% bitrate compared to end-to-end image compression networks. To the best of our knowledge, this is the first LI-based work that outperforms the SAI reordering method using VVC as the codec. Besides, our model produces higher subjective quality and geometric consistency.

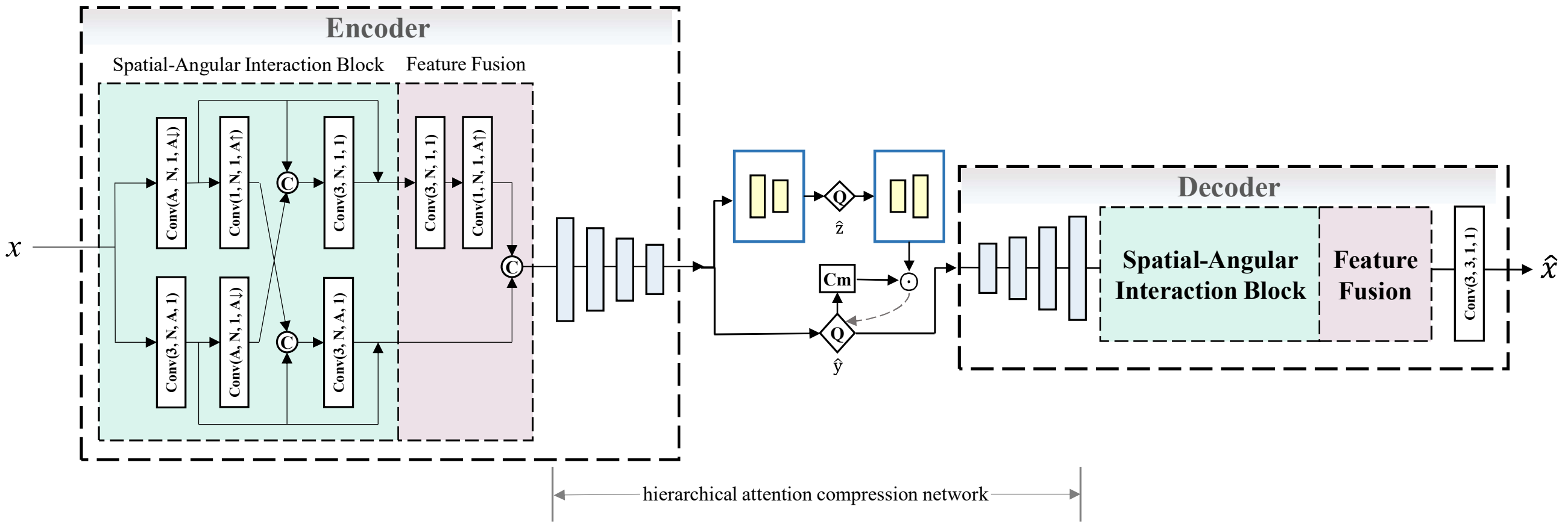

**Fig. 2.** Overview of the SADN architecture. The convolution layer is denoted as Conv (kernel size, channels, dilation, stride). A and N are 13 and 48 in our implementation respectively.

The remainder of this paper is organized as follows. Section 2 describes the proposed method and dataset. Experimental results are provided in Section 3. The conclusions are drawn in Section 4.

## 2. THE PROPOSED METHOD

In the proposed method, an LI is used as the input since the angular information is easy to be extracted. The LI is denoted by $L \in \mathbb{R}^{AH \times AW}$, where $A$ represents the angular resolution; $H$ and $W$ represent the height and width of the LI in pixel, respectively. When the Lambertian condition holds, an MI in the LI, like the green box sample shown in Fig. 1 (a), is in size of $\mathbb{R}^{A \times A}$ containing $A \times A$ angular pixels for a focused point, while each MI presents the spatial information of the LF. Based on the representation, SADN together with a large-scale LI dataset are proposed to compress LF efficiently.

### 2.1 The architecture of the proposed SADN

SADN takes the LIs as the input and reconstructs the lossy LIs at the decoder side. Its architecture is shown in Fig. 2. SADN is composed of spatial-angular interaction, feature fusion, and an image compression backbone network. Spatial-angular interaction decouples the angular and spatial information in the LI. Spatial-angular fusion fuses the decoupled information for feature extraction. The hierarchical attention compression network with large receptive field is adopted as the backbone [24] to analyze and compress the spatial-angular feature maps and synthesize the coarse LI feature maps.

**Spatial-angular Interaction.** Based on the distinctive pixel distribution, a spatial feature extractor (SFE), 3 × 3 convolution with dilation of $A$, is utilized to extract the spatial information. An $A \times A$ convolution with the stride of $A$, called angular feature extractor (AFE), extracts the angular information.

The spatial and angular information in the LI can be decoupled by SFE and AFE as

$$\mathcal{F}_{S0} = \mathrm{SFE}(L),\ \mathcal{F}_{A0} = \mathrm{AFE}(L), \tag{1}$$

where $\mathcal{F}_{S0}$ and $\mathcal{F}_{A0}$ represents the initial spatial and angular feature maps, respectively.

Since objects may be depth varying or occluded in the natural world, pixels of the unfocused point cannot map at the same MI. To address this problem, spatial-angular interaction makes the spatial and angular information fully interactive and decouples the information:

$$\begin{aligned} \mathcal{F}_S &= \mathcal{F}_{S0} + \mathrm{SFE}(\mathcal{F}_{S0}, \mathcal{F}_{A0} \times A) \\ \mathcal{F}_A &= Conv(\mathcal{F}_{A0}, \mathrm{AFE}(\mathcal{F}_{S0})) + \mathcal{F}_{A0} \end{aligned}, \tag{2}$$

where $\mathcal{F}_S$, $\mathcal{F}_A$ represent the spatial and angular feature maps. $Conv(.)$, $\times A$, and + represent convolution, up-sampling operation and skip connection, respectively. The skip connection ensures that the interaction retains the spatial and angular information of focused points, and learns the latent information with residual learning. The up-sampled angular information concatenated with $\mathcal{F}_{S0}$ is extracted by SFE to guide the spatial information extraction. The extracted spatial information by AFE is concatenated with $\mathcal{F}_{A0}$ to obtain the latent angular information.

**Feature Fusion.** In order to consider both spatial and angular information jointly, feature fusion fuses decoupled

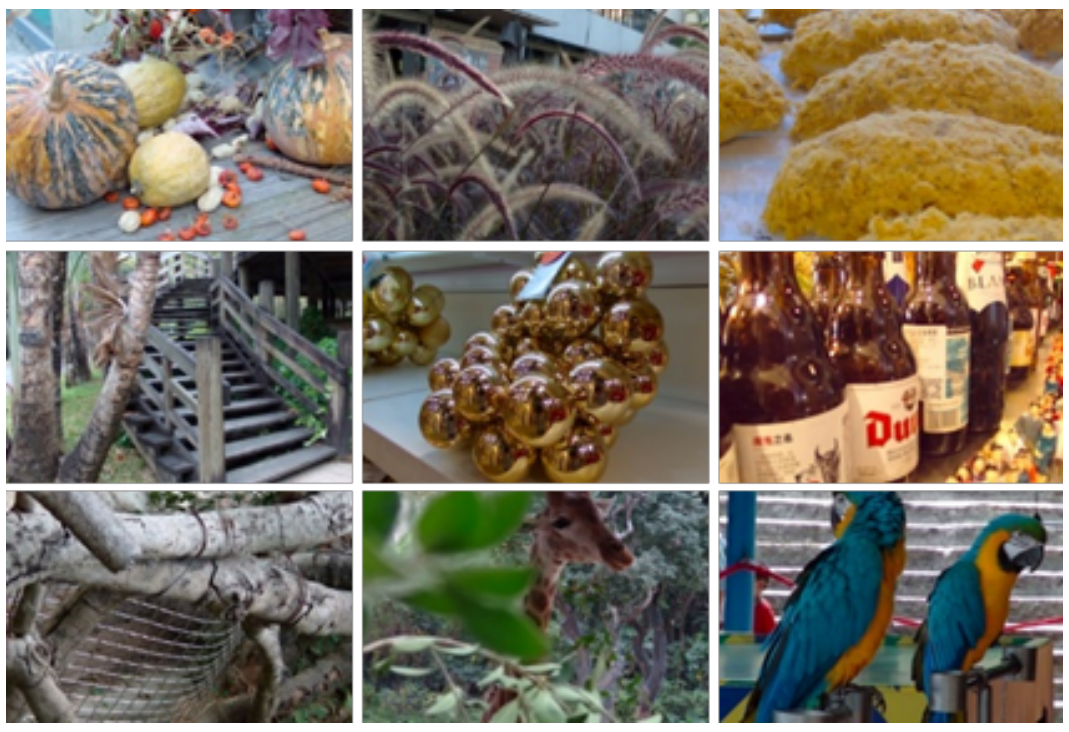

Fig. 3. Some central SAIs of LF images in the dataset.

spatial and angular information by

$$\mathcal{F}_f = \mathrm{SFE}(\mathcal{F}_S, Conv(\mathcal{F}_A) \times A), \tag{3}$$

where $\mathcal{F}_f$ denotes fused spatial-angular feature maps. Thus, analysis network with a large receptive field can fully compress $\mathcal{F}_f$ considering angular and spatial information jointly.

The loss function of the RD optimization is defined by

$$\mathcal{L} = \mathcal{R} + \lambda \mathcal{D}, \tag{4}$$

where $\mathcal{R}$, $\lambda$, $\mathcal{D}$ represent rate measured by bits per pixel (bpp), Lagrangian multiplier and Mean Square Error, respectively. Different bit rates are determined by different $\lambda$ values.

### 2.2 LF dataset

We propose a new LF image database called "PINet"[1] inheriting the hierarchical structure from WordNet [25]. It consists of 7549 LIs captured by Lytro Illum [26], which is much larger than the existing databases. The images are manually annotated to 178 categories according to WordNet, such as cat, camel, bottle, fans, etc. The registered depth maps are also provided. Each image is generated by processing the raw LI from the camera by Light Field Toolbox v0.4 [27] for demosaicing and devignetting. Some central SAIs of LF images are shown in Fig. 3 Over 34% categories have more than 20 images, and 19 categories have more than 100 images. The large-scale natural dataset benefits us to train a stable and robust algorithm.

## 3. Experimental Results

To demonstrate the effectiveness of our proposed algorithm, experiments and comparisons are conducted. The RD result, geometric consistency and qualitative results are analyzed.

### 3.1 Experiment setup

We tested the algorithms on the commonly used ICME 2016 Grand challenge test dataset with 12 LF images [28]. We compare our work with state-of-the-art (SOTA) methods GCC [14], GPR [11], SOP [16], SPR [17], and end-to-end image compression methods Minnen's [23] and Cheng's [24]. SOP and SPR are performed on HM 16.9 [29] and VTM 10.0 [30], the official test models of H.265/HEVC and H.266/VVC. Minnen's, Cheng's and our model are trained on the non-overlapping 43099 LI patches of "PINet". Due to significant distortion of the edge perspective, LF images with central angular resolution $13 \times 13$ are used for the test.

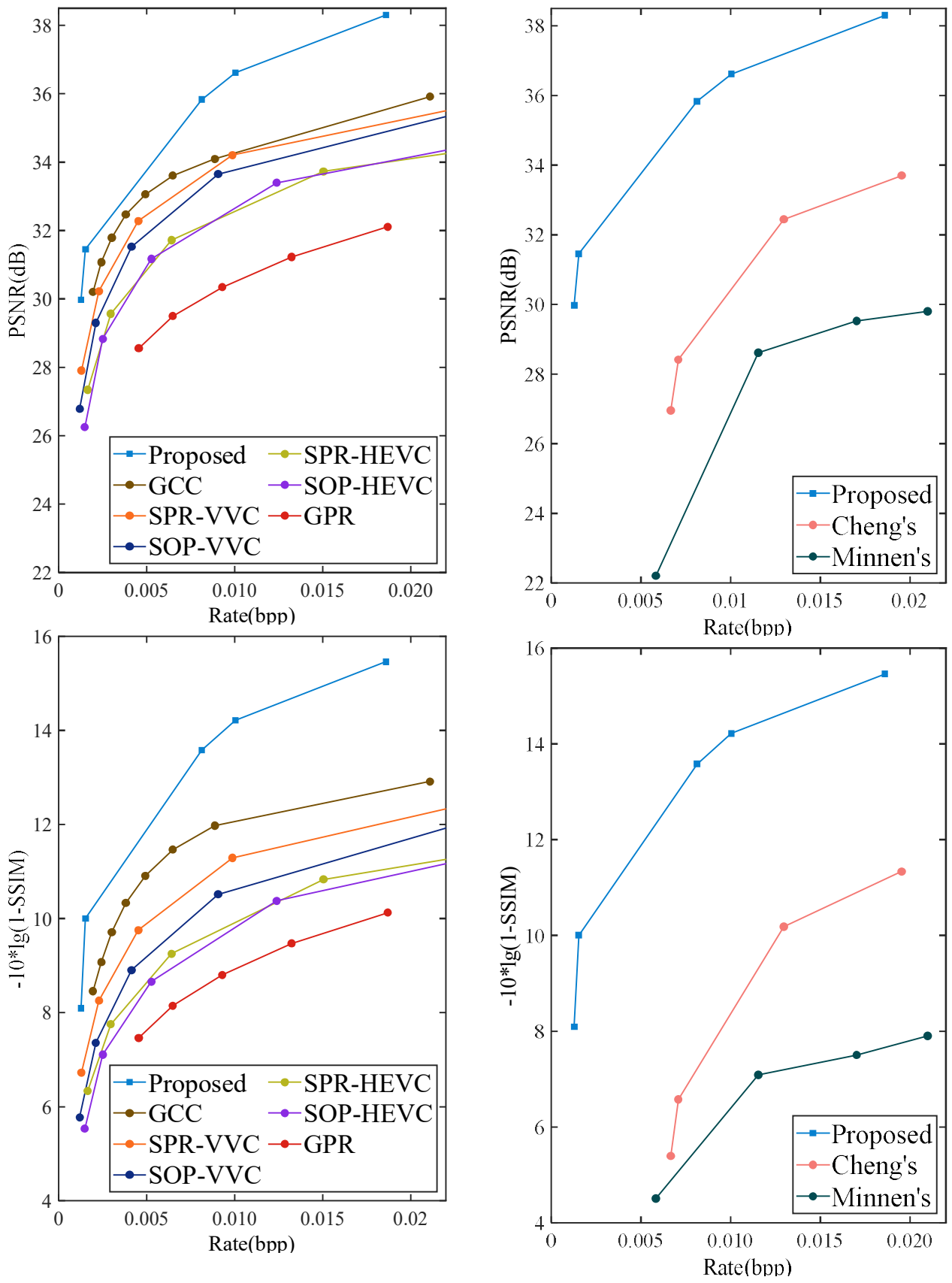

(a) (b)

Fig.4. Performance evaluation on ICME light field challenge test dataset. (a) RD curves of LF methods (b) RD curves of end-to-end methods.

### 3.2 Rate-distortion performance

The RD performance is compared in Fig. 4. In terms of PSNR and SSIM, our method achieves the best coding performance at all the test RD points with a remarkable gain compared to LF methods and end-to-end methods. Compared to the SOTA sparse sampling method GCC, our method further exploits the redundancy of LF and yields much attractive RD performance at the relatively high bitrates (>0.008 bpp). To the best of our knowledge, our method is the first LI-based work to achieve better performance than the SAI-based methods applied in VVC. The big performance gap between our method and two end-to-end image compression methods, shown in Fig. 4 (b), demonstrates the superiority of decoupling and compressing the spatial-angular information jointly.

[1] https://cloud.tsinghua.edu.cn/d/d47ad68552ec408eac94/

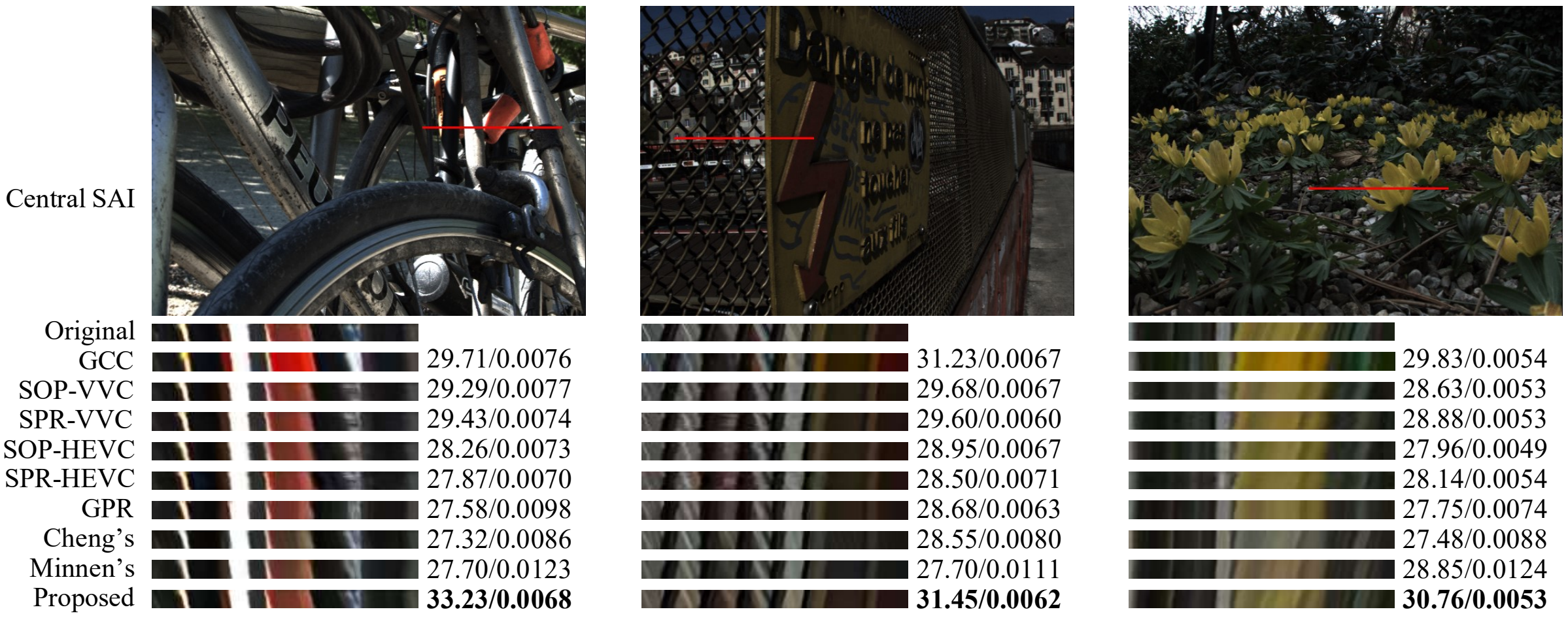

**Fig. 5.** Comparison of the EPI consistency of depth varying regions of Bikes (I01), Danger_de_Mort (I02), Flowers (I03). The quality of each image is measured by PSNR (dB)/bpp.

The BD-rate [31] performance of our method with the SOTA methods is listed in Table 1. The proposed method outperforms the other methods with considerable gain. It achieves 36.2% and 64.9% bitrate saving relative to GCC and SAI reordering methods respectively. Moreover, it outperforms the SOTA end-to-end hierarchical attention image compression networks by a 79.6% bitrate reduction. Surprisingly, SADN provides 6.579 times higher compression efficiency compared to GPR, the SOTA LI-based method.

**Table 1.** The BD-rate performance of SADN vs. LF compression methods on 12 LF images

| Method | GCC | SPR-VVC | SOP-VVC | SPR-HEVC | SOP-HEVC | GPR | Cheng's |
|---|---|---|---|---|---|---|---|
| BD-BR | -36.2% | -46.4% | -53.2% | -64.9% | -64.6% | -84.8% | -79.6% |

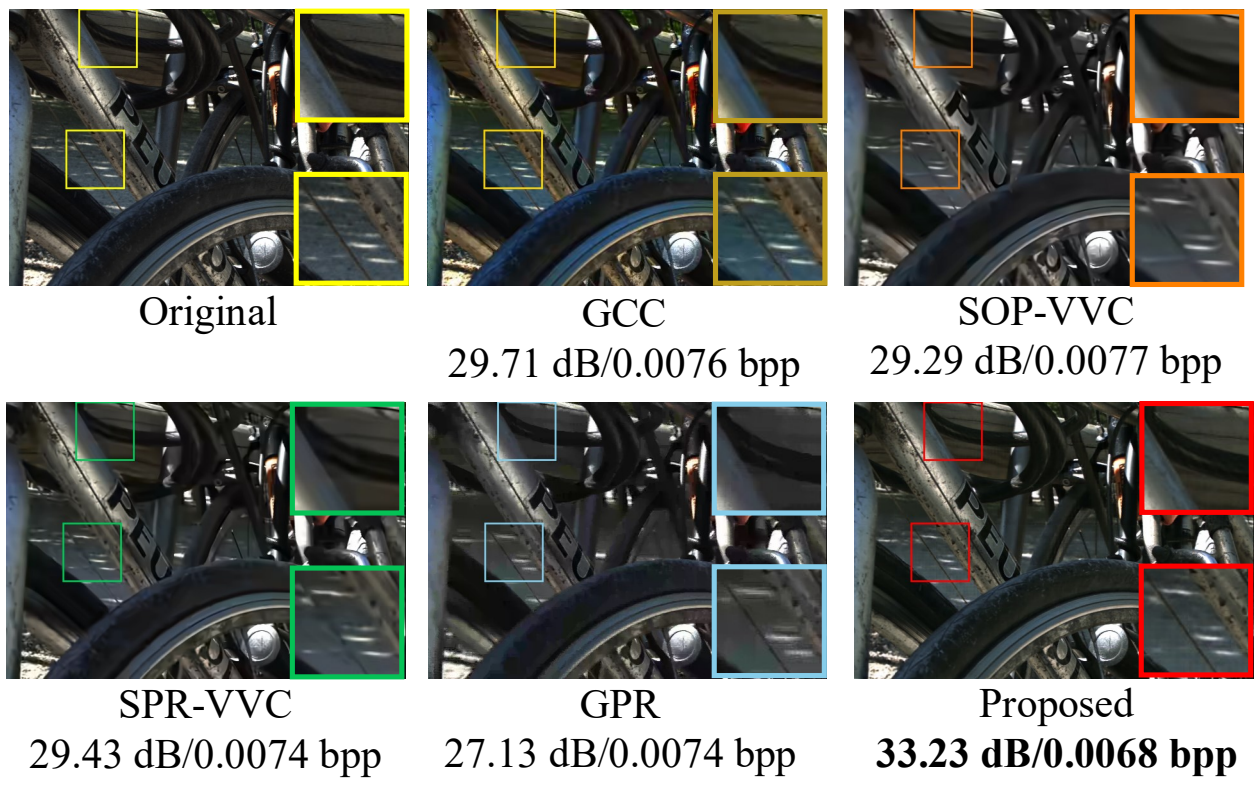

**Fig. 6.** Visualization of reconstructed central SAI of Bikes (I01) with approximately 0.0068bpp.

### 3.2 Geometric consistency and qualitative results

The Epipolar Plane Images (EPI), which contains depth varying and occluded information of the natural objects, can reflect the geometric consistency of LF. EPI consistency of depth varying regions extracted from the decoded results are shown in Fig. 5. The SAI reordering methods, SPR and SOP, show obvious distortion on EPI due to the content discontinuity of SAIs. GCC synthesizes the unsampled SAIs based on the disparity maps and fails to preserve the consistency of occluded areas. The GPR, Minnen's and Cheng's methods concentrate on the local MIs and generate smooth EPI. Our method preserves the EPI consistency remarkably at a low bitrate.

The central SAI of Bikes is extracted with approximately 0.0068 bpp at the compression ratio 3500:1 in Fig. 6. Our method generates more visually pleasing results with clear edges in bike cables. While, color distortion and blurring are easy to be observed in the results of the other methods.

## 4. CONCLUSIONS

This paper proposed a SADN to take advantage of both spatial and angular consistency into LF compression. We proposed spatial-angular interaction to decouple the spatial and angular information in LI by dilation convolution and stride convolution. Feature fusion fuses the spatial and angular information to compress the information jointly. A large-scale LF dataset consisting of 7549 images with a variety of contents and textures is built to train our model. Experimental results have demonstrated the superiority of our proposed method compared to the SOTA methods. The proposed method saved 64.9% bitrate relative to H.265/HEVC inter coding and achieved 36.2% bitrate reduction compared to SOTA sparse sampling method. It also outperforms the state-of-the-art end-to-end hierarchical attention image compression networks by 79.6% bitrate saving. Besides, the subjective quality and geometric consistency of our model outperform existing methods.

## 5. ACKNOWLEDGEMENTS

This work was supported in part by National Natural Science Foundation of China (NSFC) under grant 62131011 and Shenzhen project JCYJ20200109142810146.

## 6. REFERENCES

[1] K. Mishiba, "Fast depth estimation for light field cameras," IEEE Transactions on Image Processing, vol. 29, pp. 4232-4242, 2020.

[2] D. G. Dansereau, B. Girod, and G. Wetzstein, "LiFF: Light field features in scale and depth," in Proceedings of the IEEE/CVF Conference on Computer Vision and Pattern Recognition, 2019, pp. 8042-8051.

[3] M. Broxton et al., "DeepView Immersive Light Field Video," in ACM SIGGRAPH 2020 Immersive Pavilion, 2020, pp. 1-2.

[4] T. Ebrahimi, "JPEG PLENO abstract and executive summary," ISO/IEC JTC, vol. 1, 2015.

[5] N0122, "Exploration Experiments and Common Test Conditions for Dense Light Fields," ISO/IEC JTC 1/SC 29/WG 04, 2021.

[6] R. Ng, M. Levoy, M. Brédif, G. Duval, M. Horowitz, and P. Hanrahan, "Light field photography with a hand-held plenoptic camera," Stanford University, 2005.

[7] "Raytrix." https://raytrix.de/ (accessed.

[8] C. W. X. J. G. Bang, "m57367: Conversion between Plenoptic 1.0 Lenslet image and Multiview Images," ISO/IEC JTC 1/SC 29/WG 4, 2021.

[9] Y. Wang, L. Wang, J. Yang, W. An, J. Yu, and Y. Guo, "Spatial-angular interaction for light field image super-resolution," in European Conference on Computer Vision, 2020: Springer, pp. 290-308.

[10] T. Zhong, X. Jin, L. Li, and Q. Dai, "Light field image compression using depth-based CNN in intra prediction," in ICASSP 2019-2019 IEEE International Conference on Acoustics, Speech and Signal Processing (ICASSP), 2019: IEEE, pp. 8564-8567.

[11] D. Liu, P. An, R. Ma, W. Zhan, X. Huang, and A. A. Yahya, "Content-based light field image compression method with Gaussian process regression," IEEE Transactions on Multimedia, vol. 22, no. 4, pp. 846-859, 2019.

[12] C. Conti, L. D. Soares, and P. Nunes, "Light field coding with field-of-view scalability and exemplar-based interlayer prediction," IEEE Transactions on Multimedia, vol. 20, no. 11, pp. 2905-2920, 2018.

[13] C. Conti, L. D. Soares, and P. Nunes, "HEVC-based 3D holoscopic video coding using self-similarity compensated prediction," Signal Processing: Image Communication, vol. 42, pp. 59-78, 2016.

[14] X. Huang, P. An, Y. Chen, D. Liu, and L. Shen, "Low Bitrate Light Field Compression with Geometry and Content Consistency," IEEE Transactions on Multimedia, pp. 1-14, 2020.

[15] P. Astola and I. Tabus, "Wasp: Hierarchical warping, merging, and sparse prediction for light field image compression," in 2018 7th European Workshop on Visual Information Processing (EUVIP), 2018: IEEE, pp. 1-6.

[16] F. Dai, J. Zhang, Y. Ma, and Y. Zhang, "Lenselet image compression scheme based on subaperture images streaming," in 2015 IEEE International Conference on Image Processing (ICIP), 2015: IEEE, pp. 4733-4737.

[17] A. Vieira, H. Duarte, C. Perra, L. Tavora, and P. Assuncao, "Data formats for high efficiency coding of lytro-illum light fields," in 2015 international conference on image processing theory, tools and applications (IPTA), 2015: IEEE, pp. 494-497.

[18] D. Liu, L. Wang, L. Li, Z. Xiong, F. Wu, and W. Zeng, "Pseudo-sequence-based light field image compression," in 2016 IEEE International Conference on Multimedia & Expo Workshops (ICMEW), 2016: IEEE, pp. 1-4.

[19] N. Bakir, W. Hamidouche, O. Déforges, K. Samrouth, and M. Khalil, "Light field image compression based on convolutional neural networks and linear approximation," in 2018 25th IEEE International Conference on Image Processing (ICIP), 2018: IEEE, pp. 1128-1132.

[20] Z. Zhao, S. Wang, C. Jia, X. Zhang, S. Ma, and J. Yang, "Light field image compression based on deep learning," in 2018 IEEE International Conference on Multimedia and Expo (ICME), 2018: IEEE, pp. 1-6.

[21] T. Zhong, X. Jin, and K. Tong, "3D-CNN Autoencoder for Plenoptic Image Compression," in 2020 IEEE International Conference on Visual Communications and Image Processing (VCIP), 2020: IEEE, pp. 209-212.

[22] J. Ballé, D. Minnen, S. Singh, S. J. Hwang, and N. Johnston, "Variational image compression with a scale hyperprior," presented at the International Conference on Learning Representations (ICLR), 2018.

[23] D. Minnen, J. Ballé, and G. Toderici, "Joint autoregressive and hierarchical priors for learned image compression," arXiv preprint arXiv:1809.02736, pp. 1-22, 2018.

[24] Z. Cheng, H. Sun, M. Takeuchi, and J. Katto, "Learned image compression with discretized gaussian mixture likelihoods and attention modules," in Proceedings of the IEEE/CVF Conference on Computer Vision and Pattern Recognition, 2020, pp. 7939-7948.

[25] G. A. Miller, "WordNet: a lexical database for English," Communications of the ACM, vol. 38, no. 11, pp. 39-41, 1995.

[26] "Lytro Illum." https://www.lytro.com/. (accessed.

[27] D. G. Dansereau, O. Pizarro, and S. B. Williams, "Decoding, calibration and rectification for lenselet-based plenoptic cameras," in Proceedings of the IEEE conference on computer vision and pattern recognition, 2013, pp. 1027-1034.

[28] M. Rerabek, T. Bruylants, T. Ebrahimi, F. Pereira, and P. Schelkens, "Icme 2016 grand challenge: Light-field image compression," Call for proposals and evaluation procedure, 2016.

[29] "HEVC Official Test Model." https://vcgit.hhi.fraunhofer.de/jvet/HM/-/tags (accessed.

[30] "VVC Official Test Model " https://vcgit.hhi.fraunhofer.de/jvet/VVCSoftware_VTM/ (accessed.

[31] G. Bjontegaard, "Calculation of average PSNR differences between RD-curves," VCEG-M33, 2001.